# Optimal Algorithms for Near-Hitless Network Restoration via Diversity Coding

Serhat Nazim Avci and Ender Ayanoglu
Center for Pervasive Communications and Computing
Department of Electrical Engineering and Computer Science
University of California, Irvine
Irvine, CA 92697-2625

*Abstract*—Diversity coding is a network restoration technique which offers near-hitless restoration, while other state-of-the-art techniques are significantly slower. Furthermore, the extra spare capacity requirement of diversity coding is competitive with the others. Previously, we developed heuristic algorithms to employ diversity coding structures in networks with arbitrary topology. This paper presents two algorithms to solve the network design problems using diversity coding in an optimal manner. The first technique pre-provisions static traffic whereas the second technique carries out the dynamic provisioning of the traffic on-demand. In both cases, diversity coding results in smaller restoration time, simpler synchronization, and much reduced signaling complexity than the existing techniques in the literature. A Mixed Integer Programming (MIP) formulation and an algorithm based on Integer Linear Programming (ILP) are developed for pre-provisioning and dynamic provisioning, respectively. Simulation results indicate that diversity coding has significantly higher restoration speed than Shared Path Protection (SPP) and *p*-cycle techniques. It requires more extra capacity than the *p*-cycle technique and SPP. However, the increase in the total capacity is negligible compared to the increase in the restoration speed.

## I. INTRODUCTION

In wide area networks, cable cuts occur approximately 4.39 times a year per 1000 sheath miles [1]. The fact that networks fail frequently is a motivation for network designers to develop restoration methods for efficient and effective recovery. Network failures occur in many different forms, but single link failures are the most common form. They consist of 70% of all the failures [2], which leads us to focus on single link failures in this paper, although our techniques can be generalized to multiple failures relatively easily. There are many single link failure recovery techniques and each offers a different tradeoff in terms of different performance metrics. The prominent restoration techniques can be listed as ring-based restoration, mesh-based restoration, and the *p*-cycle technique [3], [4], [5]. Two main metrics to evaluate the performance of a restoration technique are the capacity efficiency and the restoration speed. For the telephone network, the industry goal has been to achieve a restoration time less than 50 ms. If the disruption lasts less than 50 ms humans do not perceive the failure; whereas for more than 50 ms, they do. In networking circles, hitless switching is considered to be the ultimate restoration technique, whereby restoration time is much less than 50 ms, as close to zero as possible [3]. For IP traffic, it is always desirable to decrease the restoration time to smaller values due to the complexities introduced by the different layers of the networking hierarchy.

The so-called 1+1 and 1:1 Automatic Protection Switching (APS) techniques are the simplest examples of mesh-based restoration techniques. They reserve a link-disjoint backup path for every primary path, active and inactive respectively. They are not implemented due to their extremely high spare capacity requirements. These schemes are extended to $M:N$ and $M+N$ APS, which require $N$ link-disjoint backup paths to protect the $M$ primary paths from any $N$ link failures. Today, most commonly, mesh-based restoration techniques are grouped into two, namely path-based restoration and link-based restoration. The mesh-based protection schemes that employ sharing of the spare capacity among different primary paths offer high capacity efficiency. However, the disadvantage of these techniques are lower restoration speed and higher signaling complexity.

In the 1990s, when high-speed optical transmission standards for long-haul networks were being developed, an accompanying restoration technique was incorporated into one standard. This standard is known as Synchronous Optical Network (SONET). The concept is based on protection switching over reserved capacity of multi-node ring structures, known as self-healing rings or SONET rings. More than 100% capacity is deployed over self-healing rings to match and protect all of the affected traffic over the failed links. However, due to geographical deployment of self-healing rings, required redundant capacity in fiber miles exceeds 100%.

A technique that combines the speed advantage of SONET rings and the capacity efficiency of mesh-based restoration is known as the *p*-cycle protection [6]. The speed advantage of the *p*-cycle protection comes from the elimination of most of the cross-connect configurations that are required to reroute the traffic after the failure. "Hot-standby" [7] and "pre-cross-connected trials" (PXT) [8] are developed based on the same idea.

Treating the link failure recovery problem in the context of the erasure channel model was introduced in [9], [10] and was called diversity coding. The technique provides automatic protection against single link failures and eliminates the complex and time consuming signaling and rerouting operations after the failure. Diversity coding is also able to save capacity by coding multiple flows over the same link. As a result, both the restoration time and spare capacity reduction goals can be achieved, albeit within certain limits. $N$ primary links are

This work is partially supported by NSF under Grant No. 0917176.

protected using a separate $N+1^{st}$ protection link which carries the modulo-2 sum of the data signals in each of the primary links. If all of the $N + 1$ links are disjoint, in other words physically diverse, then any single link failure can affect only one of them and the decoder will extract the failed data by applying modulo-2 sum over the received links. In [4], [11], diversity coding was applied to arbitrary network topologies, using a heuristic algorithm. There, it is shown that diversity coding is much faster than a typical Shared Path Protection (SPP) technique known as source rerouting, and the $p$-cycle technique. When the destination node for the primary paths is common, diversity coding can provide near-hitless restoration.

In [12], instead of mapping the diversity coding structures into arbitrary topologies, we introduced a technique which inputs a solution of SPP [13] and converts the sharing structure into a coding structure. This makes the restoration automatic, faster, and simpler with some slight extra capacity. This technique is called Coded Path Protection (CPP). CPP provides, in addition to coding inside the network, decoding inside the network as has been sought for within the context of network coding. The poison-antidote analogy [14] should be referred to in order to understand the encoding and decoding structures inside the CPP.

This paper can be decomposed into two parts. The first part addresses the pre-provisioning problem of the static traffic whereas the second part deals with the dynamic traffic.

## II. DIVERSITY CODING TREE

In [4] and [11], a heuristic algorithm was employed to pre-provision the static traffic by diversity coding. However, that algorithm is suboptimal and has high complexity. An optimal algorithm for that purpose can be realized with a Linear Programming (LP) formulation. In this paper, we present an optimal algorithm in a Mixed Integer Programming (MIP) formulation. The algorithm in which static traffic demands are routed and protected optimally using diversity coding is called *Diversity Coding Tree* [15]. This approach maps diversity coding to networks with arbitrary topology and static traffic scenarios. As a side advantage, this algorithm is simpler and more scalable than the heuristic algorithm.

The optimal diversity coding tree algorithm uses ideas from a *p*-cycle approach that is based on a cycle exclusion technique [16]. In [16], *p*-cycles are generated and placed in the network concurrently without prior enumeration. The diversity coding tree algorithm builds trees in order to generate primary and protection paths for the connections that share the same destination node. Each diversity coding tree structure has two separate trees forming the primary and protection paths of the connections, respectively. The primary tree produces the primary paths of the connections, which are protected by the same tree. In this tree, there is a link-disjoint path from each source node to the destination node. The protection tree serves as the common protection path for all of the connections protected by the same tree. The branches of this tree can merge until they reach the root of the tree, or the destination node. The primary tree and the protection tree of the same diversity coding tree structure are link-disjoint. An example is provided in Fig. 1. In this figure, nodes $S1$, $S2$, and $S3$ transmit their data, given as $a$, $b$, and $c$, respectively, to the common destination node $D$. The primary tree is shown with solid black lines and arrows. The protection tree is disjoint to all these links used by the primary tree and it encodes the data to be protected on the tree structure, shown in this case with dashed black lines and arrows.

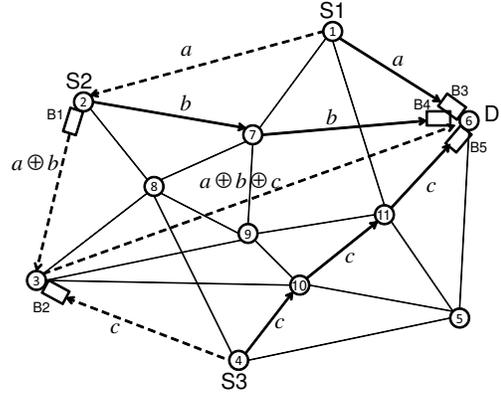

Fig. 1. An example of the diversity coding tree structure.

It is observed that when the connection demands in the same coding group have a common destination node, then it brings two advantages. First, it reduces the complexity of the LP substantially by decomposing the problem into smaller subproblems without loss of optimality. Every subproblem inputs only the set of connections with a specific destination node. Therefore, connections are divided into groups based on their destination nodes and each group is run independently. Second, it eliminates the feedback links between the destination nodes and the decoding nodes used in the general diversity coding approach. As a result, the complexity of the LP formulation, the restoration time, signaling, and synchronization complexity is reduced. The solution is applicable to any static traffic requirements, within size limits.

The boxes $B1$, $B2$, $B3$, $B4$, and $B5$ in Fig. 1 are buffers that synchronize the primary tree and the protection tree for near-hitless switching. In order to achieve near-hitless switching, each path in the same coding group is required to arrive to the destination node at the same time instant. If so, restoration can be carried out within a sub-ms time interval, which includes failure detection and node processing time. The delay values of the buffers can be calculated with the help of a variable

- $d^i_{x,y,z,v}$ : Total time when signal $i$ traverses from node $x$ to node $v$ over intermediate nodes $y$ and $z$.

Then buffer delays are calculated according to the example in Fig. 1, assuming $d^a_{1,2,3,6}$ is the longest path in the coding group and $d^a_{1,2,3} \geq d^c_{4,3}$

$$B1 = d^a_{1,2} \tag{1}$$

$$B2 = d^a_{1,2,3} - d^c_{4,3} \tag{2}$$

$$B3 = d^a_{1,2,3,6} - d^a_{1,6} \tag{3}$$

$$B4 = d^a_{1,2,3,6} - d^b_{2,7,6} \tag{4}$$

$$B5 = d^a_{1,2,3,6} - d^c_{4,10,11,6}. \tag{5}$$

The MIP formulation is provided below. The input param-

eters are provided as

- $G(V, E)$ : Network graph,
- $S$ : The set of spans in the network,
- $N$ : Enumerated list of all connections,
- $a_e$ : Cost associated with link $e$,
- $T$ : Maximum number of diversity coding trees allowed, typically one third of the number of connections in each subproblem,
- $\Gamma_i(v)$ : The set of incoming links of each node $v$,
- $\Gamma_o(v)$ : The set of outgoing links of node $v$,
- $\alpha$ : A constant employed in the algorithm, chosen very small,
- $\beta$ : A constant employed in the algorithm, chosen very large.
- $t$ : Diversity coding tree index
- $i$ : Connection demand index
- $s_i$ : Source node of the connection demand $i$

Next we provide the variables. With one exception, they are binary and take the value of 0 or 1.

- $n(i,t)$ : Equals 1 iff connection $i$ is routed and protected by the diversity coding tree $t$,
- $d_e(t)$ : Equals 1 iff the primary tree of $t$ passes through link $e$,
- $c_e(t)$ : Equals 1 iff the protection tree of $t$ passes through link $e$,
- $p_v(t)$ : A continuous variable between 0 and 1, resulting in an MIP formulation. It keeps the "voltage" value of node $v$ in the protection tree of $t$. It is possible to set this variable as an integer larger than 0 but that makes the simulation slower.
- $g_v(t)$ : Same as $p_v(t)$ except it is used for the primary tree of $t$.

The structure of the primary tree and the protection tree differ depending on the node they traverse. They have a different behavior in the destination nodes than the rest of the nodes. The following equations build the primary tree of each diversity coding structure. If the node $v$ is a destination node, then

$$\sum_{e \in \Gamma_i(v)} d_e(t) = \sum_{i=1}^{N} n(i,t) \quad \forall e \in E, t, \qquad (6)$$

$$\sum_{e \in \Gamma_o(v)} d_e(t) = 0 \quad \forall e \in E, t. \qquad (7)$$

If node $v$ is not a destination node in the primary tree, the equations are modified as

$$\sum_{e \in \Gamma_o(v)} d_e(t) = \sum_{i=1, s_i=v}^{N} n(i,t) + \sum_{e \in \Gamma_i(v)} d_e(t) \quad \forall e \in E, t. \qquad (8)$$

The following inequalities build a valid protection tree for each diversity coding tree structure. If the node $v$ is a destination node, the protection tree must terminate at the destination node using one of the incoming links. The outgoing links of the destination node should not have any tree branches. To ensure these properties, the required inequalities are

$$\sum_{e \in \Gamma_i(v)} c_e(t) \geq \frac{\sum_{i=1}^{N} n(i,t)}{\beta} \quad \forall e \in E, t, \qquad (9)$$

$$\sum_{e \in \Gamma_o(v)} c_e(t) \leq 0 \quad \forall e \in E, t. \qquad (10)$$

If it is a source node then there must be one outgoing link that belongs to the diversity coding tree. At the intermediate nodes, if at least one branch of diversity coding tree gets into the node then this node must forward the tree to the destination (root) node over one of its outgoing links. The inequalities needed under these rules are

$$\sum_{e \in \Gamma_o(v)} c_e(t) \geq \frac{\sum_{i=1, s_i=v}^{N} n(i,t)}{\beta} + \frac{\sum_{e \in \Gamma_i(v)} c_e(t)}{\beta} \quad \forall e \in E, t. \qquad (11)$$

In order to prevent getting cyclic (or loop) structures inside the trees, we choose to assign two "voltage" values to each node in the tree, as in [16], for the primary and protection tree respectively. These are not actual voltage values, instead they are each a metric used as a variable in the formulation. A detailed explanation can be found in [15].

The inequality below expresses what is described in the previous paragraph mathematically for the primary tree.

$$g_v(t) - g_u(t) \geq \alpha \cdot d_e(t) - (1 - d_e(t)) \quad \forall e = u \to v, \forall t. \qquad (12)$$

The same principle is applied to the protection tree as

$$p_v(t) - p_u(t) \geq \alpha \cdot c_e(t) - (1 - c_e(t)) \quad \forall e = u \to v, \forall t. \qquad (13)$$

In addition, the formulation has a set of inequalities to satisfy the link disjointness property between the primary tree and the protection tree. This is ensured by the following constraint

$$d_e(t) + c_e(t) \leq 1 \quad \forall e \in S, t. \qquad (14)$$

The link-disjointness criterion between the primary paths of the same tree is satisfied while building the primary trees.

The final constraint ensures that a connection can be routed and protected by only one diversity coding tree structure

$$\sum_{i=1}^{T} n(i,t) = 1 \quad \forall i. \qquad (15)$$

The objective function is

$$\min \sum_{e \in E} \sum_{t=1}^{T} a_e(d_e(t) + c_e(t)). \qquad (16)$$

### A. Pre-Provisioning Results

In this section, we evaluate the performance of diversity coding on pre-provisioning of the static traffic. The performance metrics are the capacity efficiency and the worst-case restoration time. Capacity efficiency is the total capacity required to route and protect the connection demands. The performance of diversity coding is evaluated with SPP and *p*-cycle protection. For *p*-cycle protection, optimal cycle-exclusion based ILP for joint capacity placement (JCP) and spare capacity placement (SCP) algorithms from [16] are employed depending on the network scenario. The SCP design algorithm for SPP is taken from [13]. The algorithm of SPP is suboptimal due to the extremely high complexity of the optimal ILP formulation of SPP.

We have two test networks to analyze the comparative performance of these three techniques. These networks are the NSFNET network [17] and the Smallnet network [16]. Their topologies are given in Fig. 2 and Fig. 3, respectively. In those figures, numbers next to the nodes are node indices, whereas the numbers next to the links are costs (lengths) of using that link. In the Smallnet network, link lengths are set to 100 kilometers. The traffic matrix of the Smallnet network consists of uniformly selected 143 random demands. The traffic matrix of the NSFNET network consists of 250 random demands, which are chosen using a realistic gravity-based model [18]. Each node in the NSFNET network represents a U.S. metropolitan area and their population is proportional to the weight of each node in the connection demand selection process. In Smallnet network, JCP is carried out. In NSFNET network, since both the network size and the traffic matrix are relatively large, SCP is carried out instead of JCP. The primary paths are routed using the shortest paths and the algorithms minimize the total spare capacity. Simulation results for the NSFNET network and the Smallnet network are given in Table I and Table II, respectively. $TC$ means total capacity to route and protect the traffic. The restoration times of the *p*-cycle protection and SPP are derived from formulations in [4]. The restoration time of diversity coding is simply

$$RT = F + D,$$

where $F$ is the failure detection time and $D$ is the node processing time. The maximum value for $D$ and $F$ are taken as 300 ms and 10 ms respectively [12]. The propagation delay difference in the restoration time formulation of diversity coding is eliminated because of the buffers, which equalize the arrival time to the destination node of each path in the same coding group. The symbol $X$ refers to the configuration time of an optical cross-connect (OXC).

We know from the LP software that the best results have not been achieved for some nodes due to computer memory limitations. Therefore, there is still some space for improvement

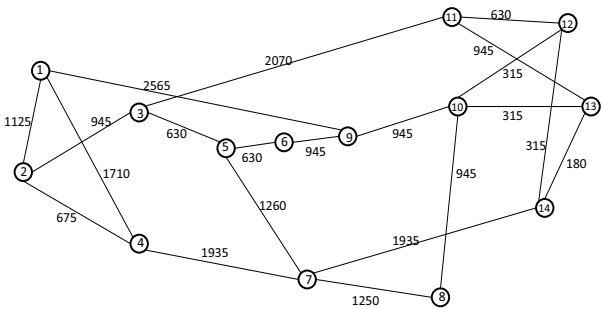

Fig. 2. NSFNET network.

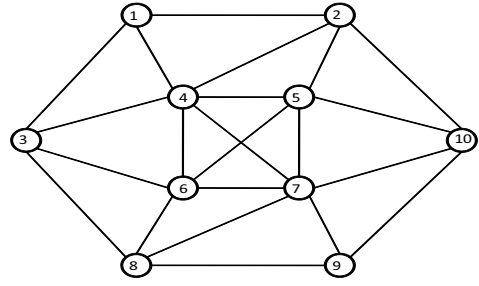

Fig. 3. Smallnet network.

TABLE I
SIMULATION RESULTS OF NSFNET NETWORK

| NSFNET Network, 14 nodes, 21 spans | | | | | |
|---|---|---|---|---|---|
| Scheme | $TC$ | $RT$ for different $X$ values (ms) | | | |
| | | 0.5ms | 1ms | 5ms | 10ms |
| Div. Cod. | 1208764 | 0.31 | 0.31 | 0.31 | 0.31 |
| SPP | 1048490 | 99.91 | 103.91 | 135.91 | 175.91 |
| *p*-cycle | 1143600 | 75.97 | 76.87 | 80.87 | 85.87 |

TABLE II
SIMULATION RESULTS OF SMALLNET NETWORK

| Smallnet Network, 10 nodes, 22 spans | | | | | |
|---|---|---|---|---|---|
| Scheme | $TC$ | $RT$ for different $X$ values (ms) | | | |
| | | 0.5ms | 1ms | 5ms | 10ms |
| Div. Cod. | 41500 | 0.31 | 0.31 | 0.31 | 0.31 |
| SPP | 34800 | 8.81 | 12.81 | 44.81 | 84.81 |
| *p*-cycle | 30200 | 8.0 | 8.5 | 12.5 | 17.5 |

in the diversity coding tree algorithm.

As seen from the results, diversity coding is much more faster than both of the techniques in each network. For pre-provisioning of the static traffic, it is less capacity efficient than SPP in both of the networks. On the other hand, the restoration speed increases on average hundred times over SPP. The restoration time of SPP increases as the expected time of OXC configuration and test increases. Realistically, in some cases it may take seconds. The *p*-cycle technique also results in higher capacity efficiency than diversity coding in both networks. The capacity efficiency of the diversity coding gets closer to the capacity efficiencies of the *p*-cycle technique and the SPP, when the SCP simulations are carried out in the NSFNET network. On the other hand, diversity coding is approximately fifty times faster than the *p*-cycle technique. The restoration time of *p*-cycle technique may increase if a link is protected via multiple *p*-cycles. In this case, end nodes of the failed link have to configure multiple OXCs simultaneously. Some nodes may not be able to carry out these configurations in parallel. Therefore, restoration time of *p*-cycles can significantly increase in some cases. It is observed that capacity efficiency of the *p*-cycle technique vanishes while going towards more sparse networks. SPP can have smaller restoration time than the *p*-cycle technique for small values of $X$ because the propagation delay around the longest *p*- cycle cancels the advantage coming from savings of the OXC configurations, especially in long-haul networks.

## III. Dynamic Provisioning

In a dynamic environment, connections are dynamically set up with a bandwidth-on-demand paradigm. Therefore the connection requests are provisioned one-by-one as they arrive. The future information about the connection demands such as in the case of pre-provisioning is not known. The dynamic provisioning of these connections against single link failures should be investigated in a different framework. This section is on the dynamic provisioning technique and the design algorithm via diversity coding.

The design algorithms for dynamic provisioning are usually heuristics since ILP based algorithms can be too complex to adapt to the changes in the environment. However, it can be observed that the complexity of the design problem is substantially simplified since the provisioning of each connection is carried out one-by-one instead of optimizing the whole set of connections at once. Combining this reality with a set of assumptions we made about the nature of the network, the result is an optimal dynamic provisioning algorithm with low complexity. The assumptions are

1. The existing connections cannot be rearranged due to QoS requirements.
2. At the beginning, the demand matrix is an empty set.
3. Centralized information about the state of the network is updated and conveyed to the nodes every time there is a change.
4. Every node is able to run the algorithm and calculate the routes.
5. Connections can be set up and terminate after some duration.

As in the previous section, diversity coding with a common destination node configuration is adopted. In other words, only the connections with the same destination node are coded together. This has the advantages of lower complexity, lower restoration time, lower signaling, and higher coding flexibility. Coding-decoding operations are simplified since the destination nodes within a coding group are the same. Different from the previous section, both the primary and protection paths are coded together, which adds up to the coding flexibility of diversity coding. One slight disadvantage of the extra coding flexibility in the diversity coding is the requirement of a simple feedforward error signaling.

The idea behind the algorithm is to provision the connection demands once they arrive by adding them to the existing valid coding groups. The additions should obey a set of rules to preserve the validity of the new coding group. The proof of decodability is carried out by induction. Assume that there is an existing coding structure whose received vector is

$$\begin{bmatrix} a_{11}x + a_{12}y + a_{13}z + a_{14}v \\ a_{21}x + a_{22}y + a_{23}z + a_{24}v \\ a_{31}x + a_{32}y + a_{33}z + a_{34}v \\ a_{41}x + a_{42}y + a_{43}z + a_{44}v \\ a_{51}x + a_{52}y + a_{53}z + a_{54}v \end{bmatrix} \quad (17)$$

where $x$, $y$, $z$, and $v$ are the coded signals and $a_{ij}$ are the binary coding parameters. The received matrix $\mathbf{A} = [a_{ij}]_{5\times 4}$

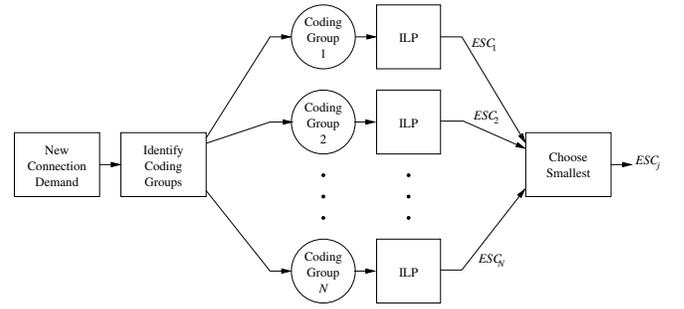

Fig. 4. Extra spare capacity is calculated for each coding scenario and the minimum is chosen.

has a "full rank + 1" property, which means if we delete one of its rows the remaining matrix would be still full rank. Deleting a row is the analogy of a single link failure whereas the full rank property assures the decodability of the signals. Therefore, it is vital to have a systematic approach to find the proper values of the variables in matrix $\mathbf{A}$. Assume that we have a new connection demand that shares the same destination node with the connections in matrix $\mathbf{A}$. The signal belonging to this connection is denoted as $l$. It is possible to enlarge the coding group by adding this new connection. There are three rules to follow to assure that this addition will not compromise the "full rank + 1" property of the new matrix

1. One of the paths of $l$ must be link-disjoint to any path in the coding group.
2. The other path of $l$ must be coded with only one path in the coding group.
3. No path in the coding group must diverge after any node.

As a result, the received matrix is transformed to the following format

$$\begin{bmatrix} a_{11}x + a_{12}y + a_{13}z + a_{14}v + 0l \\ a_{21}x + a_{22}y + a_{23}z + a_{24}v + 0l \\ a_{31}x + a_{32}y + a_{33}z + a_{34}v + 0l \\ a_{41}x + a_{42}y + a_{43}z + a_{44}v + 0l \\ a_{51}x + a_{52}y + a_{53}z + a_{54}v + l \\ 0x + 0y + 0z + 0v + l \end{bmatrix}. \quad (18)$$

The link disjoint path of the new connection is represented as the sixth row of the new matrix, as $l$ is the signal of the new connection. The other copy of signal $l$ is coded with other signals in the coding group as denoted in the fifth row. A detailed proof of how the new vector is decodable under any single link failure can be found in [15].

The $\mathbf{A}$ matrix is enlarged till it reaches its topological limit. In other words, if the three rules above cannot be satisfied within the existing topology, then a new connection group can be formed by the new connection demand itself. The $\mathbf{A}$ matrix of the new group becomes

$$\begin{bmatrix} k \\ k \end{bmatrix} \quad (19)$$

where $k$ is the signal carried by the new connection. The fact that the initial state of a new coding group satisfies the "full rank + 1" property completes the proof of decodability by induction.

We have developed an integer linear programming (ILP)

based algorithm to protect connection demands in a network. It maps the diversity coding structures into arbitrary topologies in order to protect the connections against single link failures in a cost efficient way. It is optimal under the assumptions previously stated. When the new connection is coded with one of the existing links of coding groups, there is no extra capacity incurred due to this coding operation. It leads to capacity savings and the amount of savings depends on the existing coding group topologies. The algorithm investigates each coding group and finds out how much extra spare capacity is required after adding the new connection in each coding group and selects the one which requires lowest extra spare capacity. This operation is depicted in Fig. 4. In this figure, $ESC$ means extra spare capacity required to add the new connection to each coding group. One of the coding groups in Fig. 4 is an empty group which lets the new connection demand to start a new coding group if that solution requires the lowest $ESC$. We formulated an ILP algorithm to find link disjoint primary and protection paths for every coding scenario. The cost vector of the links is adjusted for every coding scenario, depending on the topology of the coding group. This is explained in [15] with an example. Once the best available coding group is selected, the coding group topology is updated with the new connection and the algorithm is ready to incorporate a new connection demand.

The parameters of the ILP formulation to find a pair of link disjoint primary and secondary paths are as follows.

- $G(V, E)$ : Network graph,
- $S$ : The set of spans in the network,
- $N$ : Enumerated list of all connections,
- $a_e{}^1$ : Cost associated with link $e$ for the primary path,
- $a_e{}^2$ : Cost associated with link $e$ for the secondary path,
- $\Gamma_i(v)$ : The set of incoming links of each node $v$,
- $\Gamma_o(v)$ : The set of outgoing links of node $v$.

The binary ILP variables which take the value of 0 or 1 are

- $x_e$ : Equals 1 iff the primary path of the connection passes through link $e$,
- $y_e$ : Equals 1 iff the secondary path of the connection passes through link $e$.

The objective function is

$$\min \sum_{e \in E} x_e \cdot a_e{}^1 + y_e \cdot a_e{}^2. \quad (20)$$

The origination, flow, and termination of the primary path ($x_e$) and the secondary path ($y_e$) are determined by

$$\sum_{e \in \Gamma_i(v)} x_e - \sum_{e \in \Gamma_o(v)} x_e = \begin{cases} -1 & \text{if } v = s, \\ 1 & \text{if } v = d, \\ 0 & \text{otherwise,} \end{cases} \quad \forall v, \quad (21)$$

$$\sum_{e \in \Gamma_i(v)} y_e - \sum_{e \in \Gamma_o(v)} y_e = \begin{cases} -1 & \text{if } v = s, \\ 1 & \text{if } v = d, \\ 0 & \text{otherwise,} \end{cases} \quad \forall v, \quad (22)$$

The link disjointness between the primary and secondary paths is satisfied by

$$x_e + y_e \leq 1 \quad \forall e \in S. \quad (23)$$

One important note for decodability purposes is the obligation that once a primary path is coded with one of the paths in the coding group, it will stay on that path till the destination node is reached. Otherwise, the primary path may span multiple paths in the same coding group, which can impair the full rank property under some failure scenarios. This does not change the total cost since the cost of every link in the coding group is equal to zero.

### A. Limited Capacity Case

All of the discussion about the dynamic provisioning algorithm so far was based on the assumption that there is enough capacity on each link to place the primary and secondary paths. Otherwise, some of the new connection demands may be blocked due to insufficient capacity over the trail of the candidate paths. When the link capacities are limited, the cost vector of the links is adjusted within the consideration of both free capacity of each link and the topologies of the suitable coding groups. At the end, a new demand is blocked if it cannot be routed and protected with a finite cost.

### B. Connection Teardown

We need to update the algorithm when a connection leaves the demand matrix after a duration. The teardown operation consists of three steps.

- First, the connection is dropped from its coding group and the topology of that coding group is updated. This is done by subtracting the links which purely carry the signal of interest from that coding group topology. The links that carry the coded version of this signal are kept in the coding group topology.
- In the second step, the received matrix of the coding group is updated by excluding the signal associated with the leaving connection.
- In the last step, the capacity of the links that are subtracted from the respective coding group are increased by 1. It should be noted that the last step is required only if the capacity of the links are limited.

### C. Dynamic Provisioning Results

In this section of the paper, simulation results of the ILP based dynamic provisioning algorithm are compared against a $p$-cycle protection algorithm given in [19] and an ILP-based algorithm for diverse routing given in [20], which is another form of SPP. We modified the ILP formulation of diverse routing to consider the available capacity on each link. The Most-Free-Routing (MFR) algorithm for $p$-cycle protection is employed from [19] since it is the dominant algorithm in that paper. The same NSFNET and Smallnet networks are used for simulations. In contrast to the pre-provisioning simulations, every link has the same limited capacity. The objective of the algorithms is to minimize the blocking probability given specific link capacities against variable dynamic traffic loads. A new connection demand is blocked if there is not sufficient capacity over the possible paths of that connection. The traffic matrices are created as they were done in the simulations of pre-provisioning.

The techniques are compared in terms of worst-case restoration time and blocking probability in Table III, Fig. 5, and Fig. 6, respectively. It should be noted that the diversity coding algorithm for dynamic provisioning is optimal in terms of minimizing total capacity. Therefore, it may produce suboptimal results in terms of blocking probability, which means the results of diversity coding can be improved. According to the simulation results, both restoration time analysis and blocking probability analysis for dynamic provisioning are very similar to the ones in pre-provisioning since the nature of the protection structures are very similar except the design algorithms are different. Therefore, diversity coding is much faster than $p$-cycle protection and diverse routing in dynamic provisioning. The requirement of feedforward signaling when both primary and protection paths are coded marginally increases the restoration time of diversity coding compared to the case of static pre-provisioning. Diversity coding has slightly higher blocking probability than $p$-cycle protection and SPP. However, the difference in terms of blocking probability is negligible compared to the speed advantage of diversity coding.

TABLE III
RESTORATION TIME RESULTS FOR DYNAMIC PROVISIONING

| Scheme | RT for different X values (ms) | | | | | |
|---|---|---|---|---|---|---|
| | Smallnet | | | NSFNET | | |
| | 0.5ms | 3ms | 10ms | 0.5ms | 3ms | 10ms |
| Div. Cod. | 0.62 | 0.62 | 0.62 | 0.62 | 0.62 | 0.62 |
| SPP | 3.61 | 15.61 | 43.61 | 64.21 | 70.21 | 138.21 |
| $p$-cycle | 8.0 | 10.5 | 17.5 | 76.21 | 78.71 | 85.71 |

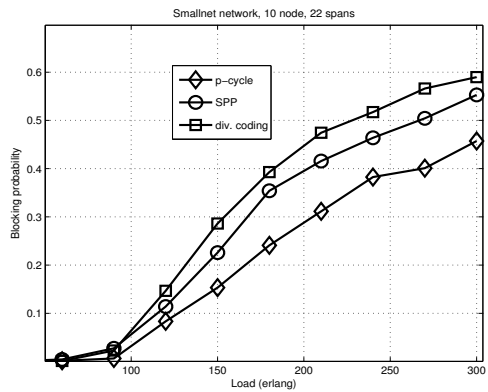

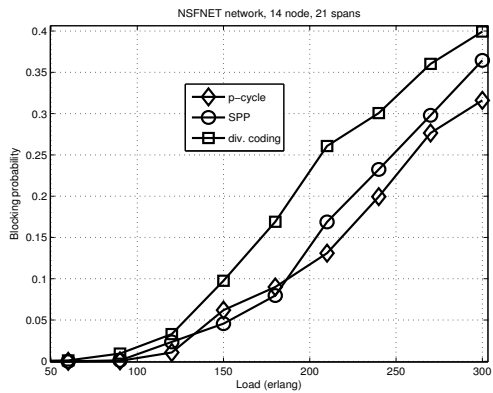

Fig. 5. Blocking probability in NSFNET network.

Fig. 6. Blocking probability in Smallnet network.

## IV. CONCLUSION

In this paper, we presented two algorithms which map the technique of diversity coding into arbitrary topologies for static and dynamic traffic respectively. The algorithm for dynamic traffic requires some assumptions to be made for optimality whereas the algorithm for static traffic is optimal without any prior assumptions. The diversity coding structure allows only connections with the same destination nodes to be coded in order to eliminate time consuming and complex operations inside the network. It also helps us to divide the problem into smaller pieces while keeping the optimality.

The first algorithm is an optimal MIP formulation to protect a set of connections by jointly optimizing the primary paths and the coded restoration paths. We call the graph that defines the optimum primary paths and the protection path the primary tree and the protection tree, respectively. These two trees form the diversity coding tree structure. A diversity coding structure both routes and protects the connections. Simulation results indicate that diversity coding is much faster than the other techniques in both scenarios. On the other hand, other restoration techniques can be more capacity efficient than diversity coding for these scenarios. However, the gap in terms of capacity efficiency is negligible compared to the gap in terms of restoration speed.

The second algorithm is dynamic provisioning of the demands that arrive one-by-one. This algorithm employs an ILP-based formulation using diversity coding to route and protect the incoming connections dynamically. That is done by adding the new connections into the suitable coding groups which have the same destination node as the new connection. The coding group which incurs the lowest extra spare capacity is chosen to add the new connection demand. If there is no available coding group, then the connection forms its own group by itself. We have carried out simulations in different networks to compare the blocking probability and restoration time performance of diversity coding, $p$-cycle protection, and diverse routing. The results are consistent with the ones in the pre-provisioning of the static traffic.